# Dating of wines with cesium-137: Fukushima's imprint

Michael S. Pravikoff (pravikof@cenbg.in2p3.fr) and Philippe Hubert (hubertp@cenbg.in2p3.fr)
*Centre d'Études Nucléaires de Bordeaux-Gradignan (CNRS/Université de Bordeaux)*
*19 Chemin du Solarium*
*CS 10120*
*33175 Gradignan cedex, France*

The well-known method of authenticating vintages by detecting $^{137}$Cs ($T_{1/2}$ = 30 years) present in the wine without opening the bottle was developed more than 20 years ago by Philippe Hubert in collaboration with the SCL of Pessac (DGCCRF). [HUB2001, HUB2009] Combined with PIXE[1] measurements performed at the ARCANE Servicesl[2] at CENBG, which tells us which elements are present in the glass of the bottle, it has become a particularly efficient tool within the PRISNA Prestations Services, particularly in the fight against fraud involving old vintages of great vintage, authentic or not.

The technique used is low background gamma spectrometry and measurements are made at the PRISNA facility. The advantage of this technique is that it does not require the opening of the bottle (a prerequisite for collector bottles) at the expense of a loss of sensitivity and a sharp increase in measurement times. With a sensitivity of the order of 0.05 Bq / l, this technique allows dating for vintage wines between 1952 and 2000, but above all it is very effective for very old vintages: indeed any bottle before 1952 does not can not contain $^{137}$Cs, even in the trace state.

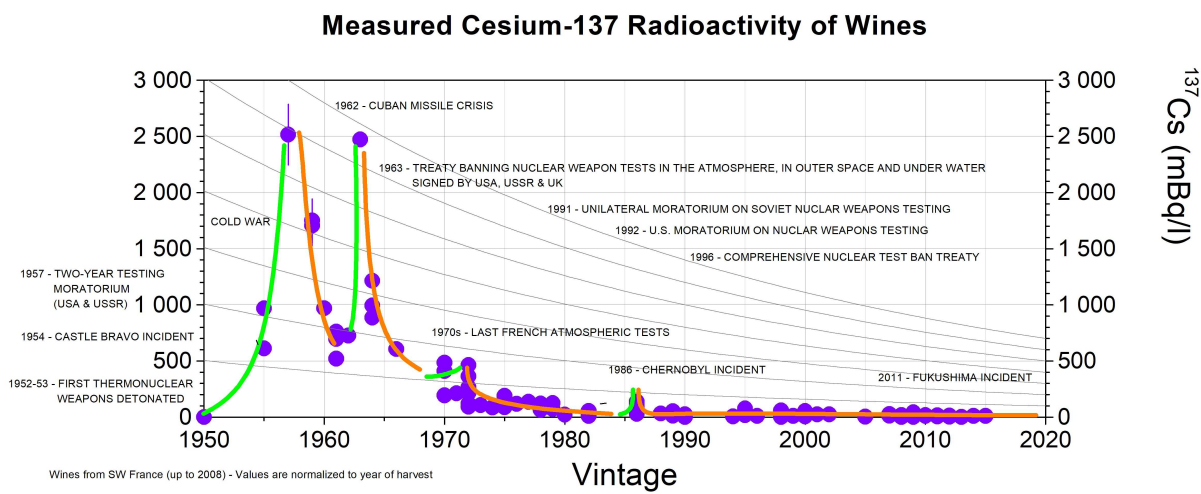

*Figure 1*

Figure 1 shows the up-to-date curve plotting the cesium-137 activity of red wines according to the vintage. As a

1  Particle-Induced X-ray Emission
2  Atelier Régional de Caractérisation par Analyse Nucléaire Élémentaire

guide for the eye, green and orange curve curves highlight periods of increasing and decreasing activity resp. The activities indicated for the measured points are those of the corresponding vintage. The light grey curves allow to correlate the activity of a given vintage with the activity measured at any given date: they simply take into account the decay period of $^{137}$Cs.

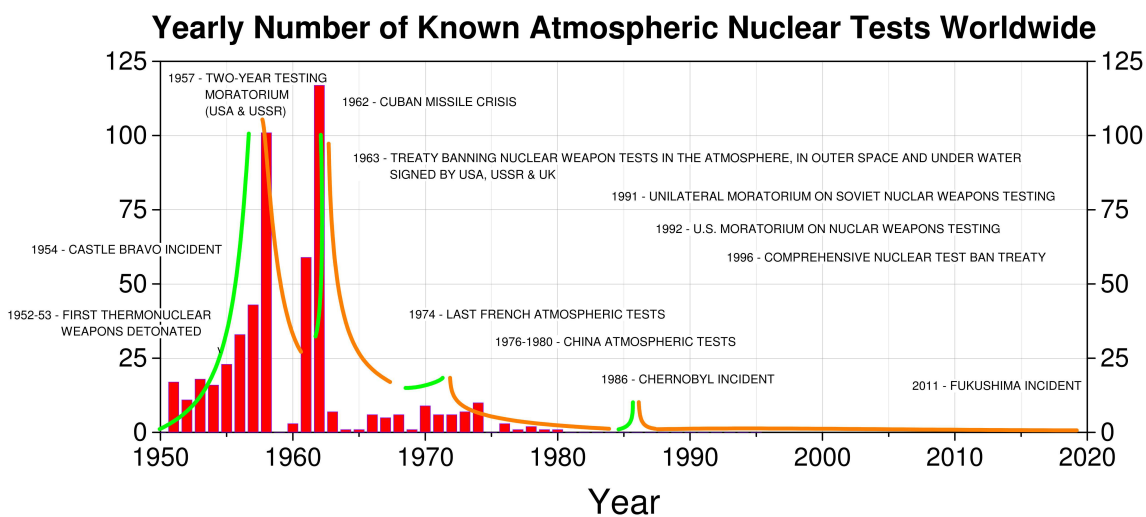

*Figure 2*

Figure 2 draws a parallel between the rise and fall of radioactivity in wines and the number of atmospheric nuclear tests known. [NIL2000] It is observed, with possibly a slight shift due to the dynamics of the movements of the atmosphere, a temporal similarity. Of course, this "correlation" has only a qualitative and not a quantitative value, if only because of the differences between all the tests, whether from the point of view of the energy released or the altitude at which the explosion took place (on the ground, at low or high altitude, or even at the limit of the space vacuum).

In January 2017, we came across a series of Californian wines (Cabernet Sauvignon) from vintage 2009 to 2012. The Fukushima incident, which took place on March 11, 2011, resulted in a radioactive cloud that has crossed the Pacific Ocean to reach the west coast of the United States. And in Northern California, there is the Nappa Valley. The idea was then to see if, as is the case in Europe following the Chernobyl accident, we could detect a variation in the cesium-137 level in these wines.

Of course, the California authorities have proceeded to countless measures in 2011 of different foods, but not wines. The values obtained always show extremely low Fukushima contamination. It is important only in seafood, such as tuna, but these are animal species from the coast of Japan.

A first series of bottles was measured according to the usual method, that is to say by placing them directly against the gamma detector low background noise without opening them. As was undoubtedly, $^{137}$Cs activities are either at the limit of detection or with a very high degree of uncertainty.

To increase the sensitivity, the solution was then to carry out a destructive analysis, namely to reduce the wines to ashes. The wine is poured into a crystallizer which is placed in an oven in which, according to a pre-established program, the temperature will gradually rise to 100° C, stay at this value for 1 hour, then rise again to 500° C, at which temperature it will remain for 8 hours. Then follow descending levels in temperature. As a general rule, the contents of a bottle (750 ml) give between 2 and 4 grams of ash. These ashes are placed in a tube adapted to the measurements by our low-noise gamma detectors of "well" type.

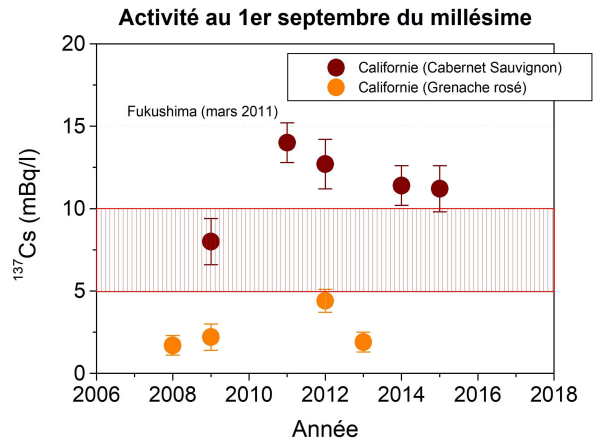

*Figure 3: California wines – the hatched area corresponds to nowadays "background" for red wines*

Figure 3 shows the first results. As was the case in France's white or rosé wine, Californian rosé bottles lead to significantly lower values than red wines. It seems there is an increase in activity in 2011 by a factor of 2. The hatched area between 5 and 10 mBq / l corresponds to the range of radioactivity values. in $^{137}$Cs measured for bottles of red wines, mainly from South-West France, dating after 1990, I.e. for which the effect of Chernobyl is no longer really quantifiable. It can be considered as the current "background noise" in $^{137}$Cs wines. Still to be verified on more samples than this was the level in California before 2011, as is the case for the 2009 wine.

The method of measurement of wines by Cesium-137 has been the subject, over the years, of several reports in the written and audiovisual press. The most recent one dates back to the beginning of March 2017. We were contacted by Dominic Byrne, documentary producer for BBC Radio 4, on the advice of Michael Egan, expert wine expert and wine detective, who dealt with various cases of Grand Cru fraud for which we brought our expertise.

The documentary entitled Wine Detectives concerns the various authentication methods and anti-wine falsification techniques, including ours. The presentation was provided by Susie Barrie, Master of Wine. The recording took place at the PRISNA platform on March 27 and was broadcast on June 28, 2017. The documentary is available for viewing on the BBC website. [BBC2017]

## BIBLIOGRAPHIC REFERENCES